\documentclass[14pt,a4paper]{article}
\usepackage{etex}
\usepackage[english]{babel}
\usepackage[utf8]{inputenc}

\usepackage{amsmath}
\usepackage{amsfonts}
\usepackage{amssymb}
\usepackage{amsthm}
\usepackage{appendix}
\usepackage{subcaption}

\usepackage{booktabs}

\usepackage{cases}
\usepackage{color}
\usepackage{comment}

\usepackage{fancyhdr}
\usepackage{float}

\usepackage[top=3cm, bottom=3cm, left=3cm , right=3cm]{geometry}

\usepackage{graphicx}
\usepackage{graphics}
\DeclareGraphicsExtensions{.jpg,.ps,.pdf,.png,.eps}
\graphicspath{{figures/}}

\usepackage{lastpage}
\usepackage{listings}
   \usepackage{multido}
\usepackage{multirow}

\usepackage[pagebackref]{hyperref}

\usepackage{pdftricks}
   \usepackage{pstricks}
   \usepackage{tikz}
\usetikzlibrary{shadows}

\usepackage{titlesec}
\usepackage{url}
\usepackage{xcolor}

\begin{psinputs}
\end{psinputs}

\def\N{{\mathcal{N}}}

\def\cX{{\mathcal{X}}}

\def\N{{\mathbb{N}}}
\def\R{{\mathbb{R}}}
\def\P{{\mathbb{P}}}

\def\1{{\mathbf{1}}}

\def\bit{\begin{itemize}}
\def\eit{\end{itemize}}

\def\bc{\begin{center}}
\def\ec{\end{center}}

\def\bcom{}

\def\edoc{\end{document}}

\usepackage[linecolor=black,textwidth=30mm,textsize= footnotesize]{todonotes} 

\title{Simulating long-term impacts of mortality shocks: learning from the cholera pandemic}
\author{Nicole El Karoui\footnote{Laboratoire de Probabilités, Statistique et Modélisation, Sorbonne-Univeristé Paris, France}, Kaouther Hadji\footnote{Skretting,  France} and Sarah Kaakai\footnote{Laboratoire Manceau de Mathématiques and Institute of Risk and Insurance, Le Mans Université, Le Mans, France.  Corresponding author: sarah.kaakai@univ-lemans.fr}}

\date{}
\begin{document}
\maketitle

{\small \paragraph{Abstract}
The aim of this paper is to study the long-term  consequence on longevity of a mortality shock. We adopt an historical and modeling approach to study how the population evolution following a mortality shock such as the COVID-19 pandemic could impact future mortality rates. In the first of  part the paper, we study the several cholera epidemics  in France and in England starting from the 1830s, and their impact on the major  development of public health at the  end of the nineteenth century. In the second part, we present the mathematical modeling of stochastic Individual-Based models.  Using the R package IBMPopSim, this flexible framework is then applied to simulate the long-term impact  of a mortality shock,  using a toy model where  nonlinear population compositional changes affect future mortality rates.}
%
\section*{Introduction}
The current COVID-19 pandemic is having a major impact on individuals and societies around the world. The invention, in one year, of several vaccines has been  a remarkable advance in medical research. However, despite these technological advances, the pandemic is disrupting the positioning of individuals in society, as public health  measures and population  management  as a whole  has become a  priority over individualized healthcare. Reconciling individuals' percpective and  global risk assessment from statistical measurements is often difficult. A lot of uncertainty is generated by short-term decisions imposed by the COVID-19 pandemic, which at the same time may have a long-term impact on future societal transformations. The pandemic has also  brought to the forefront the issue  of managing extreme mortality risks. This is not a new risk for insurers. For instance, the first excess mortality bonds were issued in 2003,  providing hedging opportunities against a short-term  increase in mortality rates (\cite{smith2017pandemic}),  and capital requirement imposed by  the Solvency II regulation includes  mortality catastrophe risk. However, the potential long-term consequences in terms of longevity of a global pandemic such as the COVID-19 have been less discussed. Yet, the major social transformations following such an extreme event could have significant impacts on future trends.   \\
In order to try and answer  this question,  it is useful to consider how societies have dealt with epidemics in the past, and what lessons can be learned from them today. 
In particular, the study of the important longevity improvements from  the beginning of  the 20th century (\cite{CUTLER2006}) has led us to analyze  the role that cholera pandemics during the nineteenth century played  on the development of massive public health transformations.  Mortality shocks induced by successive cholera outbreaks have catalyzed profound societal changes, which then had a major  impact on longevity trends in the first half of the twentieth century. The study of this historical example, initiated before the current pandemic, appears all the more relevant as long-term consequences of the COVID-19 pandemic in terms of longevity are highly uncertainty.\\
Extreme events such as pandemics question the ability of historic mortality data to be a ``good guide to the future", even in the long-term, and the ability of   extrapolatory statistical mortality models to produce accurate forecasts. We argue that limiting studies to the  mortality of older individuals constitutes  a substantial loss of information for understanding long-term impacts of mortality shocks, and that modeling and  studying the whole population evolution is a necessary  complement. \\
In the first part of this paper (Section \ref{SectionCholera}), we discuss the cholera epidemics  that took place in France and in England during the nineteenth century.  We show how the differences in the demographic, scientific, socioeconomic  and political  contexts in the two populations, may have impacted  the management of those epidemics, and study their consequences on the national and international  development of public health.  We also focus on the systematic collection of mortality data put in place in both countries, which resulted in the creation of the national institutes that we know today, and their use in the decision making process during the cholera epidemics. Finally, the section ends with a discussion on  how this historical example shed light on  contemporary issues and the need for modeling the whole population evolution.  \\
In Section \ref{SectionModelPop}, we 
present a brief overview of microsimulation models and stochastic Individual-Based-Models. The latter  provide both  a general mathematical framework and simulation methodology for modeling heterogeneous population dynamics with interactions. In particular, they can be used to model the  nonlinear dependency of individual mortality rates on the population composition.  Finally, this flexible framework is applied in Section \ref{SectionModelPop} on  a toy model example. Using the R package IBMPopSim (\cite{IBMPopSim}), we show how a mortality shock could affect the  changing composition of the population, which in turn determines future mortality trends. 
\section{The cholera pandemic as a catalyst for longevity improvements}
\label{SectionCholera}
In this section, we  illustrate the major impact of the cholera epidemics  on the development of  social  measures and public health, which later on played a critical role on the increase of the quality and duration of life.  We take on an historical viewpoint, based on the example of the spectacular cholera pandemics that occurred during the nineteenth century, in the midst of the Industrial Revolution. This historical period is particularly interesting, as  a period of major social changes, during which  frequent  and deadly epidemics had the most important social and economic consequences. 
The cholera pandemics, which struck fear and left indelible marks of blue-black dying faces on the collective imagination (Hence the nickname ``blue death''), compelled decision makers to undertake efficient public action. \\
We  focus on  cholera outbreaks in England and France, in order to illustrate the profound changes which occurred at different levels (city, state and international), which give valuable insight on contemporary challenges. Our goal is not to give a comprehensive description of the cholera pandemics, but rather to detail a number of points from the well documented experience of these two countries (with a particular focus on London and Paris), which have drawn our attention. In particular, we emphasize on how  different political and social climate in the two countries 
may have influenced the collection and use of data, as well as the development of public policies. We also focus on the  international dimension of issues raised by cholera pandemics, in the context of colonial expansion and increased global circulation, which was  materialized by the organization of more than ten international conferences.
\subsection{Cholera pandemics during the nineteenth century}
The nineteenth century was a period of major social changes, 
with the Industrial Revolution leading to a total upheaval of society, associated with unbridled urban sprawl and unsanitary living conditions.
 For instance, the population in Paris doubled from 1800 to 1850 to attain over one million inhabitants (\cite{JARDIN1983}), while London grew by 2.5 fold during those 50 years, to attain more than 2 million inhabitants (\cite{chalklin2001rise}). In 1831, the the population was of about 14 million inhabitants  in England and Wales ( \cite{ONSCensusWales})  and 33 million in France. Since then, the population has more than doubled in France and has been multiplied by almost 4 in England and Wales. \\ 
 This particular context fostered outbreaks of infectious diseases. But the two countries differ radically by their political regime: in England, Queen Victoria reigned for most of the century (1837-1901), while France experienced multiple regime
changes after the fall of Napoleon I, with: the Bourbon restoration (1814-30), the July Monarchy (1830-48), the Second Republic (1848-52) following the 1848 Revolution, the Second Empire (1852–70), and the Third Republic (1870-1940).
 
 The century was marked by six cholera pandemics, which caused millions of fatalities over the world. The first pandemic originated from India in 1817, and spread along trade and military routes to Europe in the 1830s. France and England were affected by four subsequent outbreaks (1831-32, 1848-54, 1866-67 and 1888-89). 
In England and Wales in 1831,  more than 21 thousand deaths were reported during the first outbreak (\cite{UNDERWOOD1948}),  and the second outbreak was more severe, with more than 53 than thousand fatalities. 
In France, the first outbreak caused about 100 thousand deaths  ({\cite{NoteStatistiqueCholera}), about twice the mortality rate than in England.  
The 1854 outbreak was the most important in France, with more than 140 thousand deaths from the disease, and was characterized by a wider diffusion on the territory (\cite{bourdelais1978marche}). Big cities such as London and Paris were the most affected. 
In comparison, the number of excess death assigned to COVID-19 as an underlying cause has been estimated to be around 47 000 in England and Wales and 29 000 in France  during the first wave (mid-February to end of May 2020, \cite{kontis2020magnitude}). 
In the face of these recurrent crisis, it was thus critical to undertake public health actions at different levels (city, state, international), under the pressure of sometimes conflicting economic interests and while the disease  was still poorly understood. \\
\noindent {\sc \small Scientific context} 
Cholera is an acute diarrheal illness, whose symptoms include severe watery diarrhea, vomiting or dehydration, and which can lead to death within hours if not treated. At the beginning of the nineteenth century, little was known about the disease, and the prevailing theory was the miasmatic theory, which traces its origin back to Hippocrates and predicted wrongly that the passing on of the disease was airborne. 
 The 19th century saw major development in epidemiology, medecine and mathematics, which play an increasingly important role in society. In particular, a breakthrough in the modeling of mortality was made by  the English actuary Gompertz in  1825 with his famous ``law of human mortality",  
describing age specific mortality rates 
 (\cite{KIRKWOOD2015}, \cite{GOMPERTZ1825}). 
 At the same time,  a systematic collection of mortality data is also put in place during the first half of the century, leading to the creation of national statistical institutes laying the foundation of contemporary institutions. In the following, we detail the place of data collection and analysis during the cholera epidemics which is still an important issue for understanding the COVID-19 pandemic. 
\subsection{Cholera in England}

The intensity of the first cholera outbreak in England in 1831, combined with the growing influence of advocates of public health, brought to light the need for public measures to improve sanitation. At that time, a lot of reformers considered that statistics were a prerequisite for any intervention, and the enthusiasm in the field expanded very quickly, which is somehow reminiscent of the current craze for data science. 
In this context, the General Register Office (G.R.O) was created in 1836, with the aim of centralizing vital statistics. England and Wales were divided in  2193 registration sub districts, administered by qualified registrars (often doctors). In charge of compiling data from registration districts, W. Farr served as statistical superintendent from 1839 to 1880 and became ``the architect of England's national system of vital statistics'' (\cite{eyler1973william}).\\
In his {\em Report on the Mortality of Cholera in England, 1848-49 }(\cite{FARRreport}), Farr and the G.R.O produced almost four hundred pages of statistics.
His main finding,  which has later on proven to be false, 
 was the existence of an inverse relationship between cholera mortality rates and the elevation of registration districts above the Thames, which validated his beliefs in the miasmatic theory. \\
 During the second half of the eighteenth century, the Sanitary movement flourished in England. It was actually J. Snow  who was one of the first to claim that cholera communication was waterborne in  his book \textit{On the mode of communication of cholera}, first published in 1849. Although Snow's theory was not widely accepted, he contributed to raising the issue of water quality. 
 In 1848, advisory local boards of health at city level,  under  the supervision of the General Board Health (G.B.H).  For the first time , the Metropolis Water Act of 1852 introduced regulations for private water supply companies,  although  not following all recommendations of the G.B.H (\cite{hardy1984}, \cite{millward2007distribution}).\\
At the time of the 1853-54 cholera outbreak, only one company had complied with the new regulations, and it was competing with another company drawing water from a highly polluted area.  The perfect conditions for a full-scale experiment were brought together, and Farr and Snow joined their investigations to conclude that without doubt, water played an important role in the communication of the disease.  
This famous experiment is often cited as the birth of epidemiology (\cite{brody2000map}), and the G.R.O statistics were decisive in supporting and validating Snow's theory. Yet the ambition to find causal factors by the sole analysis of data - which constitutes a major challenge in today's data science era - is not devoid of risks, and
Farr's elevation law is a textbook case of an unexpected correlation that turns out to have a great influence. \\%
In 1866, a smaller outbreak hit London, and the reintroduction of sewage contaminated water by the {East London Water Company} caused 908 deaths in just one week 
 (\cite{dupaquier1989cholera}). However, the events of 1854 and 1866 provided sufficient evidence to force the English political class to guarantee the supply of clean water. 
London's drainage system was completed in 1875, and among  measures taken were the Rivers Pollution act (1876) or the carrying out of monthly water reports (\cite{hardy1984}). \\
 However, the importance of economic interests in the management of sanitary issues should not be overlooked. 
 Despite the overwhelming amount of evidence during the 1866 outbreak, the Medical Officer himself tried to exonerate the East London Water Company. 
Interest groups in municipal councils were also most influential in passing public health measures, with sometimes conflicting interests. \cite{millward2007distribution} cites the example of Wakefield where the project of drawing clean water from a new source was abandoned due to adverse effects that this hard water could have on textile factories. These examples show the importance of the municipal level in the development of public health measures, and an important heterogeneity in decision processes depending on local circumstances. The level of granularity relevant in order to understand changes in a population is still a difficult question today.
\subsection{Cholera in France}
 It is interesting to put in perspective England's experience with cholera with France's, where the social nature of the cholera epidemics played a determinant role in the management of the disease.\\
French authorities were alerted by their consular authorities about the first pandemic 
 as soon as  1817, and  the former military A. Moreau de Jonès was asked to monitor the evolution of the disease. In a remarkable report 
(\cite{MoreaudJones1831}), he gave considerable details on the spread of the epidemic, care of the sick, and precautions to be taken to limit the expansion.  He was also  one of the first to clearly attest that cholera was incontestably contagious. In $1833$, Moreau de Jonès became the first chief of the Statistique Generale de la France (SGF), the nearest equivalent to Farr's position at  the G.R.O, and contributed to the development of Statistics and its applications in France.  Unfortunately, little attention was paid to his findings, due to the strong opposition to the idea of the disease being contagious.\\
 Indeed, many French doctors believed in the miasmatic theory, and attributed the spread of the disease to miasmas originating from infectious sites in unsanitary and crowded housings. In addition, the political class was also   opposed to declaring the disease contagious, due to quarantine measures which were seen as a barrier to trade (\cite{LeMee1998}). Despite the lack of scientific arguments, the anticontagionist theory had the most important political implication, by drawing the attention to the development of public hygiene and the fight against unsanitary living conditions.\\ 
Already in 1802, a Public Hygiene and Salubrity council was  established,  
with the aim to report on sanitary conditions in factories, cemetries or  slaughterhouses (\cite{BOYER2010}). From 1822, councils slowly spread out to the main provincial towns. However, their function was only advisory.  In 1829, the  scientific journal {\em Annales d'hygiène publique et de médecine légale} dedicated to the study of public health was founded, accompanied with a growing interest in studying the links between longevity and living conditions. An example is the pioneer work of Villermé, who  established a relationship between wealth and mortality rates, by comparing mortality data in Paris per borough with tax authorities statistics (\cite{VILLERME1830}, \cite{MIREAUX1962villerme}).\\
In the end of $1831$, the cholera epidemic was expected to reach France from England. In Paris, a particularly elaborate hierarchy of special commissions was established in 1831, but the tense political situation didn't allow for real measures to be taken. Indeed, social unrest among the lower classes during the first epidemic, who saw the disease ``as a massive assassination plot by doctors in the service of the state'', were the worst fears of the government (\cite{KUDLICK1996}) and  paralyzed decision-making.\\ 
 The outbreak was however the subject of many statistical studies, even if the data collection was not as centralized as in England. In 1832, the journal \textit{Gazette médicale de Paris} published a special issue on Cholera, including a description of the excess mortality by area, profession, housing type...  In 1834, Benoiston de Chateauneuf studied daily death statistics in Paris, and established that the cholera excess mortality was more important in unhealthy houses in poorer neighborhoods. These statistical surveys all confirmed the anticontagionist theory, despite many approximations and misinterpretations. For instance, an excess cholera mortality in one side of the Chaillot street was attributed to  the fact than one side of the street was inhabited by workers and the other side by burghers. \cite{DODIN1992} showed by reworking the data that one side of the street was actually supplied with polluted water from the Seine and not the other side (See also the analysis of the data used by Benoiston de Chateauneuf and Villermé in \cite{LECUYER2000argent}). Be that as is may, the cholera outbreak was revealing of workers living conditions, and statistics played a determining role as evidence that politicians couldn't ignore in the years that followed.\\
The year 1848, marked by the revolution,  was a turning point. The perception of the lower classes had changed, with the idea of struggling against destitution in order to prevent revolt (\cite{KUDLICK1996}).  
Contrary to the first epidemic, the reaction to the second outbreak, which reached France, at the end of the year was more peaceful. As in 1832, the disease was not declared contagious by the authorities (\cite{dupaquier1989cholera}), and  statistical work all supported the miasmatic theory. In particular, the comparison of cholera mortality rates in Paris during the first and second outbreak indicated that mortality rates had fallen in the historic center where urban development work had been realized (\cite{LeMee1998}). These studies acted as catalysts for new legislation and the  development of urban planning.\\
 In 1850, a law on unhealthy housing was passed, and municipalities were in charge of enforcing the legislation. As in England, measures taken depended heavily on local and economic interests. Health education like school hygiene was also developed in the second half of the century (\cite{BOYER2010}. In Paris, the municipal commission, relied on the cholera outbreaks statistics in order to ask for massive public work in the city. As a consequence, hygiene problems and unsanitary living conditions caused by the rapid growth of the population were addressed to in the following years by important public health measures. In 1854, Baron Haussman and the engineer Eug\`ene Belgrand designed the present Parisian sanitation system, a sewer network which was over 500 km long by 1870, and the water supply networks. The massive public work projects  led by Haussman under Napoléon III,  in less than two decades, still remains  a symbol of the modernization of Paris at the end of the nineteenth century (\cite{RAUX2014}). \\
n France, statistics have been extensively used to describe cholera outbreaks. However, rather then helping to understand the disease, they  provided justification for the (false) miasmatic theory. Paradoxically, this also contributed to the spectacular development of public health measures. However, the study of the cholera statistics, mainly focus on Paris, did not manage to prevent or reduce outbreaks. Indeed, the most important  outbreak in France occurred in 1854. If Paris was less affected than in 1832 and 1849, the third outbreak was much more widespread on the territory. This is attributed by \cite{bourdelais1978marche} to the increase in volume trade and traffic routes between 1832 and 1854, in the absence of measures against  direct human to human transmission.
\subsection{Cholera Pandemic and International Health Organization} 

The international dimension of  the problem raised by cholera, was widely publicized by The Lancet,  which published in 1831 a map on the international progress of cholera 
 (\cite{Koch2014}).
This map suggested a relation between human travel and the communication of the disease, accelerated by the industrial revolution in transport, in particular with steamships and railways. Cholera was regarded as an issue transcending national boundaries, which needed international cooperation to control it (\cite{Huber2006}).  Europe had succeeded in setting up an efficient protective system against the plague, based on ideas such as quarantine and "cordon sanitaire". But those measures were very restrictive and were seen as inefficient against cholera. Moreover, in the second half of the nineteenth century,  Western European countries were involved in competitive colonial expansion, and were rather hostile to travel restrictions, even if increased global circulation was a threat to populations. 
The opening of the Suez Canal in 1869 was an emblematic example of those changes.\\
Under the influence of French hygienist doctors, 
the first International Sanitary Conference opened  in Paris in 1851, gathering European states and Turkey. 
It was the first international cooperation on the control of global risk to human health, and so the beginning of international health diplomacy. 
It took more than ten international conferences over a period of over $50$ years to produce tangible results.
During the first five conferences, 
 the absence of clear scientific explanation on the origin of cholera prevented any agreement. 
It was only with the formal identification of the {\em V. cholerae} bacterium by R. Koch in 1883 - the bacterium had actually been isolated before by other scientists such as F.Pacini in 1854, but his work did not had a wide diffusion - and the work of L. Pasteur that infectious diseases were clearly identified and efficiently fought against.
Indeed, technological progress as evinced through disinfection machines could allow the technological implementation of new measures (\cite{Huber2006}). Furthermore, advances on germ theory ``allowed diplomats to shape better informed policies and rules'' (\cite{FIDLER2001}). \\ At the Seventh Conference (1892), the first maritime regulation treaty was adopted for ship traveling via the Suez Canal. During the ninth conference (1894), sanitary precautions were taken for pilgrims traveling to Mecca. Participants finally agreed that cholera was a waterborne disease in $1903$ during the eleventh conference.\\ 
The International Sanitary Conferences provided a forum for medical administrators and researchers to discuss not only on cholera but also on other communicable diseases, and brought about the first treaties and rules for international health governance. Ultimately, this spirit of international cooperation gave birth in 1948 to the World Health Organization, an agency of the United Nations, conceived to direct and coordinate intergovernmental health activities.
\subsection{Discussion} 
The need for cholera statistics contributed to the development of the field and the creation of national statistical institutes in France and England, which played a determining role in  the structuring of national demographic data as we know it today. 
In England,  the 
 remarkable organization of the G.R.O undoubtedly contributed to the  quality of today's England vital databases.  Across the Channel, France  may not have managed to create the same kind of centralized authority.  However,  the statistical division of the French territory, based on administrative structures that have little changed  over time, allows for a rather good understanding of long term evolution. Theses examples also highlight the critical impact of local structures (statistical units administered by doctors in England, administrative units in France) on the data structuring. 
 During the cholera outbreaks, very precise data were collected (cholera deaths by occupation, street...), but often on a local in big cities such as Paris or London.  \cite{LECUYER2000argent} also underline the difficulties faced by Villermé in order to define a good wealth indicator, or the hazardous extrapolation in Benoiston de Chateauneuf's work.  Beyond the administrative division, the political and scientific context in each countries played an critical role in selection of data which were collected, which led to important biases in  interpretations.   Statistics were actually mainly used to confirm (sometimes wrong) preexisting theories, such as Farr's elevation law,  or  to justify public policies as in France.  All these challenges are especially relevant in light of the multiple debates on the collection and analysis of COVID-19 data. More than 150 years later, it is still particularly difficult to estimate and compare mortality due to the COVID-19 pandemic, across and even within countries (\cite{kontis2020magnitude}). Thus, this  shows us the importance of being able to rely on theoretical models in order to test theories that have emerged from the study of data.  \\
Another  point is that  the conditions that made Farr and Snow's 1854 experiment  possible were quite extraordinary. Testing theories regarding the complex events of health and mortality in human communities is often nearly impossible. Furthermore, governments failed to come to an agreement during the first international conferences because of the lack of scientific explanations on the origin of cholera.
The need of theoretical arguments for public decisions to be made is still an important issue, especially when considering human health and longevity, for which no biological or medical consensus has emerged. \\ 
 The various responses to the outbreaks, across different countries, regions or cities, show the importance of taking into account the overall  population evolution and global context into account, in order to understand what could be the potential long-term impact of such extreme events. 
 Although cholera outbreaks occurred at about the same time in France and in England, they were experienced very differently owing to the different population social composition, political and scientific climates in both countries. In particular, the explosion of the London population, whose size was twice as large as that of Paris, brought about social problems on a much greater scale, which played a determining role as a catalyst of public health changes. Local governments also played an  important part in the public decision-making process,  and the various local economic interests led to a great heterogeneity in both countries in the development of local public health measures. Other factors may have influenced the development of the disease, such as the differences in  quality of  living  in English and French cities. \\
The international management of the pandemic also illustrates the striking timescales on which major decisions are taken. Even when theories are publicized, there are often important delays (one or two generations) before action is taken. For instance, even if Snow's theory was better known in 1866, and despite the development of germ theory in the early 1880s, political divergences prevented any action before 1892. The example of asbestos, which took 50 years to be banned after the exhibition of its link with cancer, shows us that these delays in public response did not diminish over time (\cite{Cicolella2010}).  \\[1mm]
In conclusion, the remarkable use of statistics contributed to a better management of the cholera, while it was only more than a century and a half later that a major breakthrough was made in the understanding of the origins of the disease, with the work of R. Colwell showing that the {\it V. cholerae} bacterium appears naturally in the environment and could lay in a dormant state under adverse conditions.  However, the collection and analysis of data have not been without difficulties and even errors, and most of the problems raised in the foregoing are still relevant.\\
 Through the example of England and France, we have seen how important it was to take into account the demographic, social and political environment in order to understand the evolution, management, and long-term consequences of the epidemics.  
 The same issues arise on even greater scales when we are concerned with the evolution of human longevity. From a modeling point of view, it appears critical to to consider the whole population evolution, as well as broader environment in which life courses are embedded, which requires a interdisciplinary approach. But several difficult questions have to be dealt with, among them: Which factors should be taken into account in the modeling? How? How do they interact together? Another important question is the issue of the most relevant scale (local, national, international) on which these factors should be considered.  The complexity of these issues  partly explains the popularity of extrapolatory mortality models rather then explanatory models. However, the radical societal, political or medical transformations that occurred in only 50 years over the second half of the nineteenth century raises the issue of  only studying mortality data  in order to understand past evolution and future trends.\\              
Finally, the cholera outbreaks contributed to the development of important public health measures, which played a major role in the reduction of infectious diseases over the first half of the twentieth century. 
The development of public health was instrumental in the process of the so-called demographic transition, and in particular in the spectacular mortality reduction that took place from the early 1900s.  %
For instance, \cite{CUTLER2005} estimated than the purification of water explained half of the mortality reduction in the US in the first third of the twentieth century. 
Public health should not be underestimated in the current age of ``degenerative and man-made diseases'' (\cite{BONGAARTS2014}). 
The increase of environmental risks constitutes one of the major challenges faced by contemporary societies, and public action in the aftermath of  current pandemic will play a central role in preventing and successfully reducing those risks (\cite{Cicolella2010}). 

\section{Modeling of the population dynamics for understanding longevity trends}
\label{SectionModelPop}
\subsection{Stochastic modeling of human populations}
If medical advances are one the main determinants of the mortality reduction that took place from the 1870s, and whose determinants have been extensively debated in the literature  (see e.g. \cite{CUTLER2006}),  public health measures, including those triggered by  cholera outbreaks, also  played a determining role in this reduction (\cite{BLOOM2016,CUTLER2005,CUTLER2006}).  The development of public health policies is strongly dependent on the population and global socioeconomic context, and on the other hand their impact  varies across socioeconomic subgroups and modify the population composition. \\
In terms of longevity dynamics, this suggests a nonlinear dependence of mortality rates on the population structure itself, which requires being able to model the population dynamics.  As for all human systems, the study of human population evolution is  intricate by the very nature of underlying mechanisms.  As we have seen with the example of cholera epidemics and their consequences, phenomena are often non-stationary, heterogeneous, and often include interactions taking place at different scales and with sometimes opposite effects. 
Due to these difficulties, producing a pertinent modeling directly at the macro-level is a complicated  task. A finer-grained description of the population evolution allows us to experiment with various mechanisms resulting in  changes in the population composition (for instance due to new public policies or changes in the socioeconomic context), and to study their impact on the population longevity. We start this section with a short review of stochastic population models, focusing on the recent mathematical and simulation framework developed for the study of stochastic individual-based models. We then present an toy example of such model in order to study the short and long-term impacts of a mortality shock. 
\subsection{Microsimulation models} With the rise of available data and computing power, so-called Microsimulation models have been developed in social sciences for the past decades. For instance, the MiCore tool have been developed as part of the European project Mic-Mac. We can also cite  the 
widely used microsimulator is SOCSIM. Used mainly by government and institutional bodies, 
they provide a simulation tool in order to address a  broad variety of questions, ranging from evaluating the impact of policy changes and demographic shocks to the study of kinship structure.  These model are mostly data-driven and their description often rely on the simulation algorithms. These features constitute an important limitation to their implementation, which often needs a considerable amount of data ({\cite{SILVERMAN2011}}). Furthermore, the complexity of microsimulation models can be significantly limited by computational costs. For instance, inter-individual interactions are often limited in microsimulation models, due to specification problems caused by data limitation or unobservable hidden processes {(\cite{zinn2017simulating})}, as well as too high computational costs and time. 
\subsection{Stochastic individual based models}
 In very different fields, advances in probability have contributed to the development of a new mathematical framework for so-called individual-based stochastic population dynamics, also called stochastic Individual-Based Models (IBMs). Initially developed in a Markovian setup in view of applications in mathematical biology and ecology (\cite{Fournier2004}, \cite{tranThese}, \cite{ferriere2009stochastic}), these models have a wide range of applications, and their extensions are particularly interesting for the study of human populations (see e.g. {\cite{theseharry}},  {\cite{boumezoued2016Thesis}}, \cite{kaakaibirth}). Stochastic IBMs allow the modeling in continuous time of populations dynamics structured by age and/or characteristics (gender, socioeconomic status, frailty...). The R package IBMPopSim (\cite{IBMPopSim}) provides simulation algorithms for simulating efficiently IBMs, which overcomes the often cited limitation of long simulation times of such models.\\
In the following, we give a brief presentation of IBMs before introducing our toy example.  As usual, the filtered probability space is denoted by $(\Omega, \{\mathcal{F}_t\}, \mathbb{P})$.  In an IBM, each individual is described at a given time $t$ by its date of birth  $\tau \in \R$, and set of characteristics $x(t) \in \cX$, with $\cX$ the space of characteristics.  The population at a given time $t$, denoted by $Z_t$,  is composed  of a collection of individuals, 
\begin{equation}
Z_t =\{ (\tau_i,x_i) ; \; i=1, \dots N_t \}, 
\end{equation}
with $N_t$ the number of individuals at time $t$. Note that individuals are usually characterized by their age $a=t-\tau$ at a given time rather than their date of birth $\tau$. However, it actually easier for mathematical modeling and simulation purpose to use the latter, which is constant over time.\\
The population evolution is modeled by different types of events occurring to individuals at random times, such as births and deaths events, changes of characteristics (swap events), or entry of individuals in the case of open populations. Each type of event modify the population composition. For instance, 
 \begin{itemize}
\item {\em A birth} event at time $t$ results in the addition of an individual of date of birth $\tau =t$ to the population. Its characteristics $X$ can be drawn according to a probability distribution $k^b(t,(\tau,x),x')\gamma(d x')$ depending on its parent's date of birth and characteristics $(\tau, x)$, and with $\gamma$ some probability distribution on $\cX$. The population size becomes $N_t = N_{t^-} + 1$. 
\item  {\em A death} event at time $t$ results in the removal of an individual  $(\tau,x)\in Z_{t^-}$ dies, the population size becomes $N_t = N_{t^-}-1$. 
\item {\em A swap} event results in the simultaneous addition and removal of an individual in the population. If the individuals characterized by $(\tau,x)$ changes of characteristics at time $t$, then it is removed from the population and replaced by $(\tau,X')$, where $X'$ can be drawn using a probability distribution $k^s(t, (\tau,x),x')\gamma(dx')$ depending on the individual's age and previous characteristics. 
\end{itemize}
The frequency of the various types events in the population are determined by individual intensities. Informally, the intensity $\lambda^e(t,(\tau,x), Z_t)$ at which an event of type $e$ can occur to an individual $(\tau,x)$  in the population $Z_t$, is defined by 
$$\P( \text{ event } e \text { occurs to an individual } (\tau,x) \in  \; ]t,t+dt] | \mathcal{F}_t) \simeq  \lambda^e(t,(\tau,x), Z_t)dt.$$
The event (individual) intensity  $\lambda^e(t,(\tau,x), Z_t)$ can depend on time (for instance when there a mortality reduction over time),  the individual's age $t-\tau$, its characteristics, but also on the population composition $Z_t$, since individuals are not independent and  their life course is impacted by the population they live in. Observe that the individual  intensity of swaping from characteristics $x$ to $x'$ is 
\begin{equation*}
\lambda^s (t,(\tau,x), Z_t) k^s(t, (\tau,x),x').
\end{equation*}
Similarly, the intensity of giving birth to an individual $(t,x')$ is $\lambda^s (t,(\tau,x), Z_t) k^s(t, (\tau,x),x')$. \\
Formally, the population  can be represented by a  random counting measure on $\R \times \mathcal{X}$,  $ \displaystyle Z_t = \sum_{i=1}^{N_t} \delta_{(\tau_i,x_i)}$. With this representation, we have for instance 
\begin{align*}
& N_t =  \int_{\R \times \mathcal{X}}  Z_t(d\tau, dx) = \sum_{i=1}^{N_t} \boldsymbol{1}_{\R \times \mathcal{X}} (\tau_i, x_i)\quad \text{ or } 
N_t([a,+\infty[) =  \int_{\R \times \cX }  \boldsymbol{1}_{[a,+\infty[}(t-\tau)Z_t(d\tau, dx),
\end{align*}
the number of individuals  of age over $a$. \\

\noindent  \textbf{Stochastic differential equation  driven by extended Poisson Measures} The pathwise representation of  the population evolution by "thinning" of Poisson measures is particularly interesting. The representation  is strongly connected with realizations of point processes as strong solutions of stochastic differential equations (SDEs) driven by extended Poisson measures. The last part of this subsection gives some insights on the population dynamics' SDE for a population with birth, death and swap events, but we refer to \cite{tranThese}, \cite{bansaye2015stochastic} or \cite{boumezoued2016Thesis} for more details. 
In order to define the SDE, a Poisson random measure is associated with each type of events $e=b$ (birth), $d$ (death) or $s$ (swap): \\
- For $e=b,s$, let $Q^e(ds, di, dx', d\theta)$ be the Poisson random measure defined  on the extended space $\R^+ \times \N^* \times \cX  \times \R^+ $, of intensity measure $ds \otimes n(di) \otimes \gamma(dx') \otimes d\theta$,  with $n(di)$ the counting measure on $\R^+$. This measure can be interpreted as a  set of random points $(S,I,X',\Theta)$  in  $\R^+ \times \N^* \times \cX  \times \R^+ $  where S is a  candidate event time, associated with a candidate individual to which the event could occur, indexed by  $I$, and a new characteristics $X'$ for the newborn ($e=b$) or the individual $I$ after the swap if $e=s$.  If the individual number $I$ is  in the population ($I \leq N_{S^-}$), then  the last coordinate $\Theta$ is used to determine if this candidate event actually occurs in the population. The event is accepted when  $\Theta \leq  \lambda^e(S,(\tau_I,x_I),Z_{S^-})k^e(S,(\tau_I,x_I),X')$.\\
- For $e=d$ death events, only candidates event times and candidate individuals are needed, and hence  the Poisson random measure $Q^d(ds, di, d\theta)$   is defined on $\R^+ \times \N^*   \times \R^+ $, of intensity measure $ds \otimes n(di)  \otimes d\theta$. The measure can be interpreted as a  set of random points $(S,I,\Theta)$  in  $\R^+ \times \N^* \times \R^+ $.  If the individual number $I$ is  in the population ($I \leq N_{S^-}$), then the individual dies if $\Theta$ is smaller than its death intensity: $\Theta \leq  \lambda^d(S,(\tau_I,x_I),Z_{S^-})$.\\
The population dynamics' SDE is then defined as follows: 
\begin{align}
\label{SDE}
Z_t & = Z_0 + \int_0^t \int_{ \cX \times \N^*\times \R^+}  \delta_{(s,x')} \boldsymbol{1}_{\{i\leq N_{t^-}\}} \boldsymbol{1}_{  \{\theta \leq\lambda^b(s,(\tau_i,x_i),Z_{s^-})k^b(s,(\tau_i,x_i),x')\}} Q^b(ds,di,dx', d\theta) \\
\nonumber & + \int_0^t \int_{ \cX \times \N^*\times \R^+}(\delta_{(\tau_i,x')}- \delta_{(\tau_i,x_i)}) \boldsymbol{1}_{\{i\leq N_{t^-}\}} \boldsymbol{1}_{\{\theta \leq \lambda^s(s,(\tau_i,x_i),Z_{s^-})k^s(s,(\tau_i,x_i),x')\}} Q^s(ds,di, dx', d\theta)\\
\nonumber & -  \int_0^t \int_{  \N^*\times \R^+} \delta_{(\tau_i,x_i)} \boldsymbol{1}_{\{i\leq N_{t^-}\}}\boldsymbol{1}_{\{\theta \leq \lambda^d(s,(\tau_i,x_i),Z_{s^-})\}} Q^d(ds,di,d\theta). 
\end{align}
In order to ensure the existence and uniqueness of a solution to \ref{SDE}, assumptions have to be made regarding the events intensity functions $\lambda^e$. See \cite{tranThese}, \cite{boumezoued2016Thesis} or \cite{kaakaibirth} for a discussion on these assumptions.  The population evolution can be simulated without approximation. However a simulation algorithm cannot be derived directly from the population SDE, since the jump times of the Poisson measures cannot be enumerated increasingly. The simulation algorithm is based implemented in IBMPopSim is based on the assumption that the events individual intensities are bounded by a linear function only depending on the size of the population.

\section{Simulating long-term impact of a mortality shock: a toy model}
\label{SectionToyEx}
Our goal is to study, in a simplified model,  the impact of a mortality shock  in the population,  for instance due to an epidemic outbreak, when the shock not only impact individuals' longevity, but also the population composition through swap events ( \cite{kaakai2019can}). Swap events can provide a way to model  compositional changes induced by public measures or variations in the general quality of life. Our aim is to study the impact on the population mortality of such events occurring at rates depending not only on the individuals characteristics but also on the population composition.  \\
We consider a population divided in two subgroups. Individuals are thus characterized by their date of birth $\tau \in \R$ and subgroup $ x \in \cX= \{1,2\}$. The first subgroup (subgroup 1)  is assumed to be less deprived, and a  have lower mortality rates than the second subgroup (subgroup  2). For simplicity, we do not make the distinction between males and females, but this can be easily included in the modeling.\\
In the following, we denote by $N_t^i = \int_{\R^+ \times \{i\}}  Z_t(d\tau, dx)$ the number of individuals in subgroup $i$, and $N_t^i([a,+\infty[) = \int_{[t-a,+\infty[ \times \{i\}} Z_t(d\tau, dx)$, the number of individuals over age $a$ in subgroup $i$. The model is implemented and simulated with the R package  IBMPopSim.

\subsection{Model without swap events}
The first model is a baseline model, in which the "epidemic outbreak"  only  impact  mortality rates. The mortality intensity of an individual $(\tau,x)$ depends on his  age $a =t-\tau$ and subgroup $x$. Furthermore, the   mortality intensity is assumed to decrease over, except during the period $[t_b,t_d[$ during which a  mortality shock occur.  Each individual $(\tau, x)$ at time $t$ has thus the mortality  intensity $\lambda^d(t, (\tau,x))$ :
\begin{equation}
\lambda^d(t, (\tau,x)) = \alpha_x  d(t, a) (1 + s(a)1_{[t_b,t_d[}(t)), \quad a=t-\tau, \, x=1,2.
\end{equation}
The function $(d(t,a))$ is a reference mortality rate. When $t\notin [t_b,t_d[$, mortality rates in subgroup $x$ are given by $( \alpha_x  d(t, a))$, with $\alpha_1 \leq \alpha_2$. Over the period $[t_b,t_d[$, mortality rates increase suddenly in each subgroup, at a rate increasing with age. For instance,  the COVID-19 pandemics has obviously more important impact on mortality at older ages (\cite{kontis2020magnitude}, \cite{islam2021excess}).\\
Birth rates are modeled by an age-dependent function $(\lambda^b(a))$. However, this function has no impact on the simulation outputs presented below. Indeed, we only show results for individuals over 40 years old, over a 40 years period. So all result concern individuals already present in the initial population. 
\begin{itemize}
\item {\em Simulation horizon:} The population evolution is simulated over a period $h=40$ years. The initial time is denoted by  $t_0=0$. 
\item  {\em Initial population:}  For each subgroup, the initial subgroup's  age pyramid is determined by the stable pyramids of each subgroup, which correspond to the long term population's age pyramid when mortality rates stay constant over time. A naive approach would be to consider a general population in which each age group is composed of the same number of individuals in Subgroup 1 and in Subgroup 2. However, this is not consistent with the demographic rates. Indeed,  there should naturally  be more individual in the less deprived Subgroup 1 in older age classes, due to mortality differences between the subgroup, which is reflected by the subgroups' age pyramids (see \cite{kaakai2019can} for more details). \\
The initial population is composed of 5 millions individuals, whose  age and subgroup are drawn according to the age pyramid's distribution. Figure \ref{InitialPop} shows the initial population age pyramid. Each bar (red and blue added) represent the number of individual in the age class, and the number  of individuals in each subgroup. 
 \begin{figure}[H]
\centering
\includegraphics[scale=0.5]{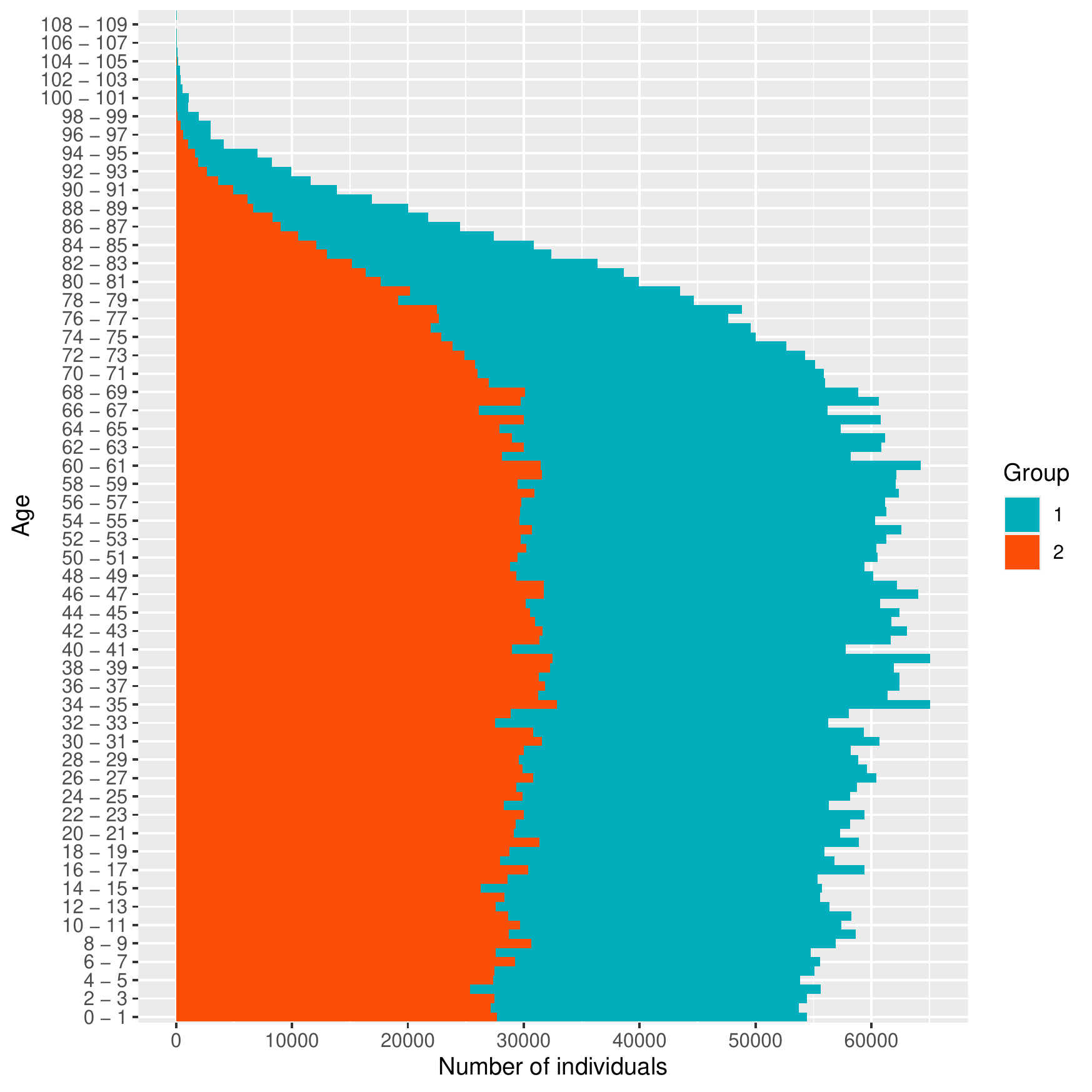}
\caption{Initial age pyramid.}
\label{InitialPop}
\end{figure}
\item   {\em Death intensities:} The reference death intensity $(d(a,t))$ is given by a forecasted Lee-Carter model using the R package StMoMo (\cite{JSSv084i03}). The model is initially fitted using one year age-specific deaths and exposures for England and Wales,   from age 0 to 100 and period  1961-2016. 
In all results, the parameters of the subgroup specific effect are  $\alpha_1 = 0.8$ and $\alpha_2= 1.3$. The mortality shock occurs from year $t_b=2$ to $t_d=5$. The magnitude of the shock is given by $s(a)=0.05 \boldsymbol{1}_{\{a<65\}} + 0.2 \boldsymbol{1}_{\{a\geq 65\}}$ (older individuals are more impacted by the mortality increase). 
\end{itemize}

\noindent \textbf{Results} The package IBMPopSim provides functions in order to compute the number of deaths by year and single year of age, as well as an exact computation of central exposures to risk.  Figure \ref{mxnoswap} represents the evolution of the estimated central death  rates $m_{at} = \dfrac{D_{at}}{E_{at}}$ for each year and age from 55 to 100. First years are in blue while last years are in green, and central death rates corresponding to $t\in [t_b, t_d[=[2,5[$ are displayed in red. For ages above 65, central death rates during the mortality shock are greater than death rates at the initial time.  Figure \ref{mx_pop12_noswap_40} represents central deaths rates at $t=40$ years for the general population and each subgroup. Central death rates are closer to death rates of Subgroup 1 for older ages, due to the more important proportion of individuals in this subgroup due to differences in mortality intensity (see Figure \ref{compoagegrpnoswap}). 

\begin{figure}[H]
\centering
\includegraphics[scale=0.4]{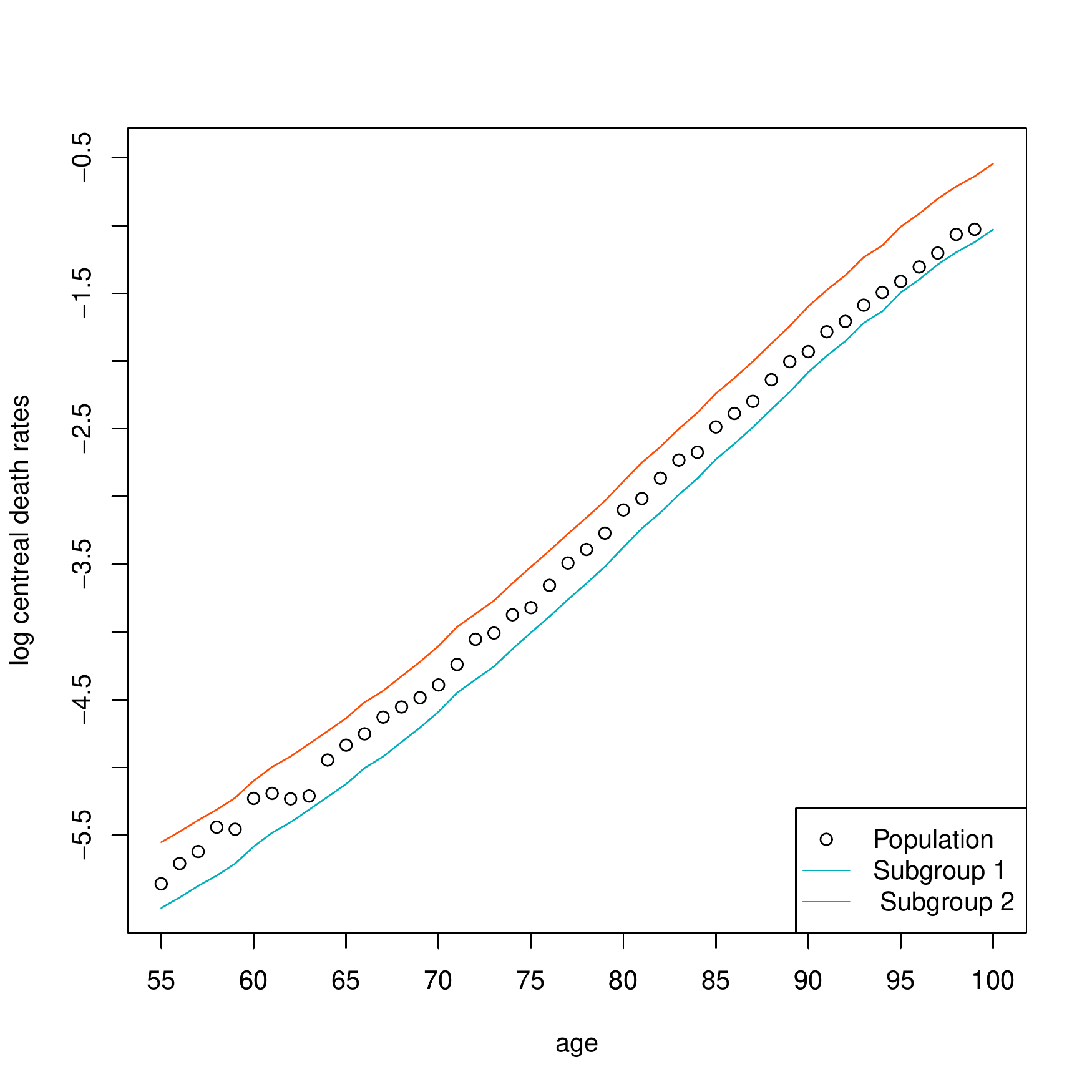}
\caption{  Central death rates for general population (black), Subgroup 1 (blue) and Subgroup 2 (red) at $t=40$.}
\label{mx_pop12_noswap_40}
\end{figure}

\subsection{ Model with impact of mortality shocks on the population} In the second model, we now take into account the short and long term effects of mortality shocks on the population, by including  compositional changes following the shock. Compositional changes are modeled by swap events (individuals can change of subgroup). First, shortly after the mortality increase,  individuals older than 40 can move from the more favorable subgroup 1 to least favorable subgroup 2. This may be caused by restricted health care access due to the mortality shock impacting older individuals, or more generally short-term decrease in the general quality of life. The magnitude of this phenomenon depend on the general population, and is assume to increase with  the number of individuals in the least favorable subgroup  2. Hence, these "adverse" swaps amplifies the negative fallout of the mortality shock. In a the second time, potential long-term positive impact of the mortality shock are modeled. We assume that on the long term, individuals can move from Subgroup 2 to Subgroup 1, following for instance an increase in the general quality of life due to important public measures, as it was the case for cholera epidemics in France and England.  This mechanism is also amplified by the number of individual in Subgroup 2. This could be the case when the number of individuals in  Subgroup 2 has an impact on the " public pressure" to take action to increase the quality of life of the most deprived. \\
Since the characteristics space $\cX=\{1,2\}$, we can define two intensity functions :$\lambda^{s_{12}}$ is the intensity of swap events from Subgroup 1 to Subgroup 2, and $\lambda^{s_{21}}$ is the intensity of swap events from Subgroup 2 to Subgroup 1.   The swap intensities are thus defined as follow:
\begin{align}
\label{Swap12}
& \lambda^{s_{12}}(t,(\tau,x), Z_t) = k_{12}N_t^2([40,+\infty[)1_{\{a \geq 40\}} 1_{[t_b^1,t_d^1[}(t) 1_{\{x=1\}}, \quad a=t-\tau.\\
\label{Swap21}
& \lambda^{s_{21}}(t,(\tau,x), Z_t) = k_{21}N_t^2 1_{[t_b^2,t_d^2[}(t)1_{\{x=2\}}
\end{align}

\begin{figure}[H]
\centering
\includegraphics[scale=0.35]{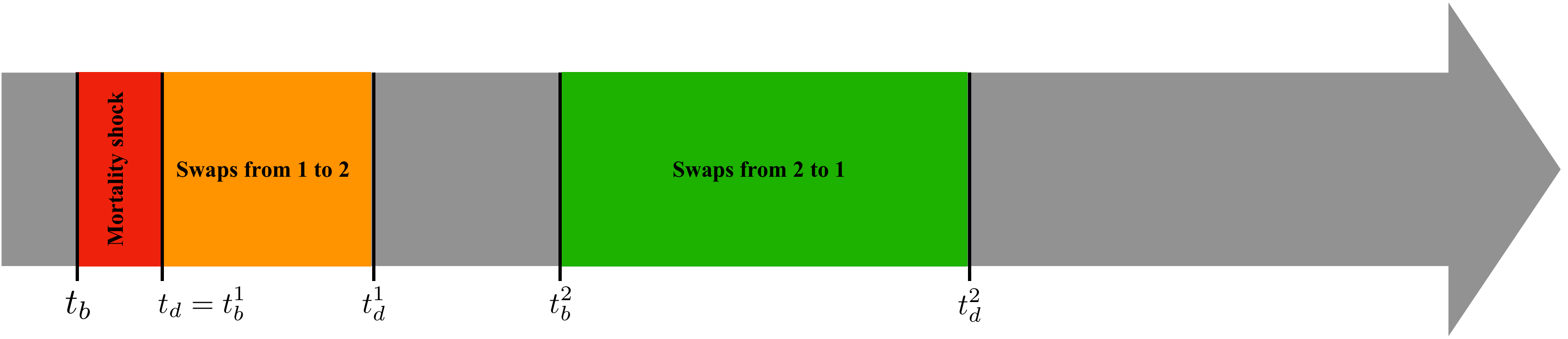}
\caption{ Timeline of events in the population}
\end{figure}
The intensity of swap events in \eqref{Swap12}-\eqref{Swap21} not only depend on the age and subgroup of individuals, but also the population composition through $N_t^2([a,+\infty[)$ and $N_t^2$, in order to take into accounts interactions between individuals. The functional form for $\lambda^{s_{12}}$ is similar to the infection rate from susceptible to infected in stochastic SIR models used in epidemiology. The intensity $\lambda^{s_{21}}$ of swap events from Subgroup 2 to Subgroup 1 differ, since it depend on the number of individuals in the subgroup of origin. \\
In the following, $k_{12} = 10^{-7}$, $k_{21} = 3\times10^{-8}$, $t_b^1 = 5$, $t_d^1=10$, $t_b^2=15$ and $t_d^2 = 25$.\\

\noindent  \textbf{Results} Figure \ref{mxtime} represents the evolution of the estimated central death rates in the general population simulated over 40 years  for each year and age from 55 to 100, and in the two models. The initial population is the same in the two models, and the first five years correspond to two simulations of the same model since swap events only occur in the second model from year $t_b^1=5$. In Figure \ref{mx2swap},  the gap between death rates (red) during the mortality shock ($[2,5[$)\ and years following is reduced, due to the adverse compositional changes occurring from year 5 to year 10. On the other hand,  the decrease in mortality rates is faster from year $t_b^2=15$ (blue-green), due the positive  changes  in the population as a long-term effect of the mortality shock. In the presence of evolution in the population composition over time, interpreting longevity trends by studying mortality data can be quite difficult. Hence, the study of the population itself  is critical. 
\begin{figure}[H]
\hspace{-1.7cm}
\begin{subfigure}[t]{.6\textwidth}
\centering
\includegraphics[scale=0.4]{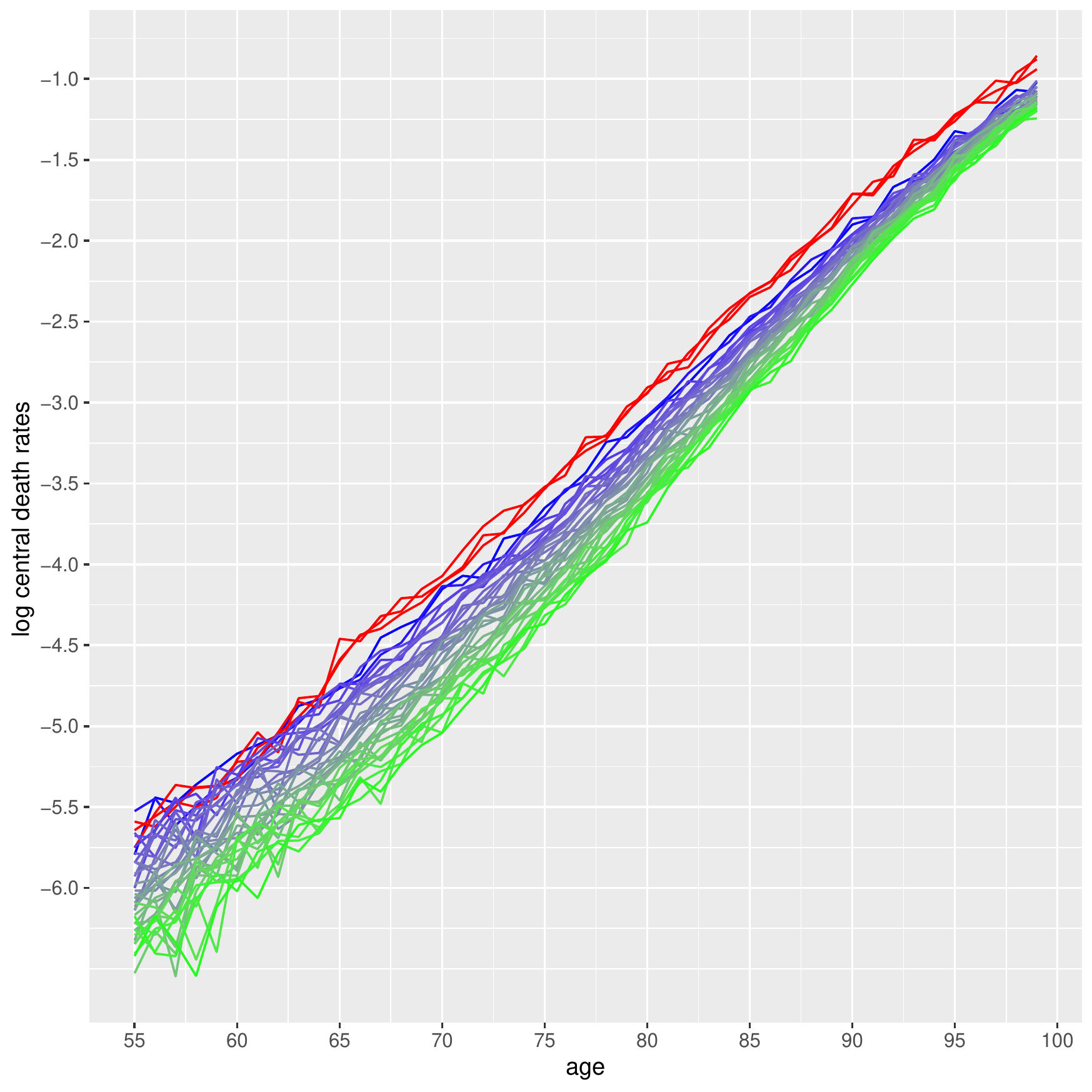}
\caption{  Model 1 (no swaps)}
\label{mxnoswap}
\end{subfigure}
\begin{subfigure}[t]{.6\textwidth}
\centering
\includegraphics[scale=0.4]{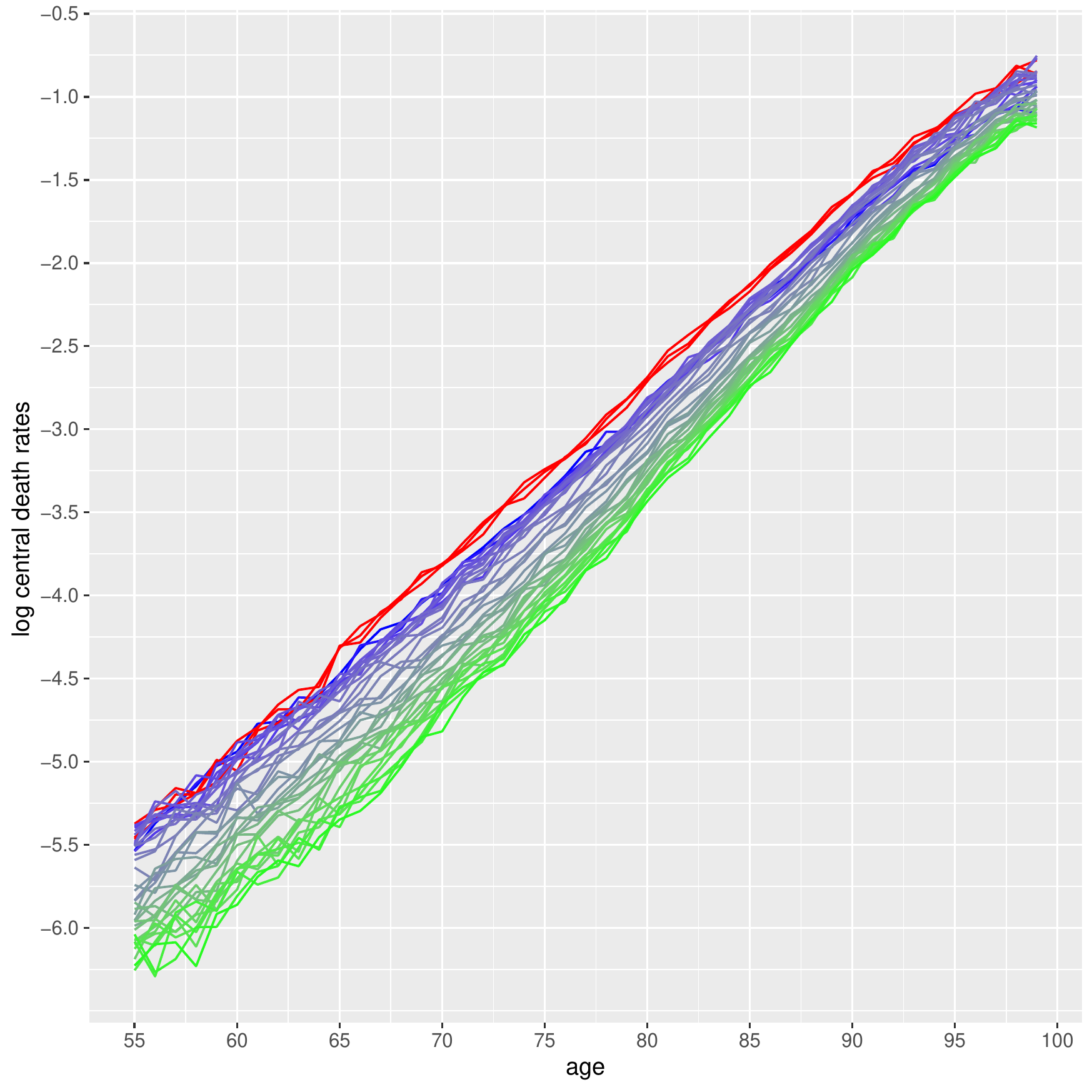}
\caption{ Model 2 (with swaps)}
\label{mx2swap}
\end{subfigure}
\caption{ Central death rates (general population) for  each year (blue to green). Red for $t\in [t_b,t_d[$).}
\label{mxtime}
\end{figure}

\noindent Figure \ref{compoagrgrp} represents the evolution over time of the proportion of individual in each subgroups, for different age groups. When there are no swap events (Figure \ref{compoagegrpnoswap}), the proportion of individuals in each subgroup stays approximately the same in each age group over time, due to the choice of the initial population.\footnote{The slight reduction of the gap between the proportion of individuals in Subgroup 1 and the number of individuals in Subgroup 2 is due to the reduction of the mortality gap between the two subgroups over time.} Figure \ref{compoagegrpswap} shows the impact of swap events on the population composition, which varies across age groups and whose impact on mortality rates is thus significantly more complicated to interpret. First, from year $t_b^1=5$ to year $t_d^1 = 10$, swap events from Subgroup 1 to Subgroup 2  cause an increase in the proportion of individuals in the (more deprived) Subgroup 2 in all age groups.  \\
Between time $t_d^1 = 10$ and $t_b^2=15$, no swap events occur in the population. However, note that the proportion of individuals in Subgroup 2 still decreases in the age class 40-50. Indeed, individuals of age 40-50 at time $t_d^1$  being in the age group 35-45 at time $t_b^1$, and thus this cohort was not fully impacted by the negative consequences of the mortality shock in terms of population composition (swaps from 1 to 2 only concerns individuals who are over 40). Even less individuals were impact by swap from 1 to 2 in the cohorts following (individuals age 40-50 at times in $[t_d^1,t_b^2[$, which causes the decrease in the number of individuals in Subgroup 2 during this period. \\
From year $t_b^2=15$ to year $t_d^2 = 25$, "positive" swap events from Subgroup 2 to Subgroup 1 (for all individuals) generate an increase in in the proportion of individuals in this less deprived subgroup,  in all age groups. This decrease is non linear, since the swap intensity $\lambda^{s_21}$ decreases with the number of individuals in Subgroup 2. For ages below 60, this decrease is amplified by the fact that younger cohorts were only partially impacted by  adverse swap  events from Subgroup 1 to Subgroup 2 after the mortality shock. \\
After time $t_d^2=25$, we observe the same mechanisms for ages above 50.  Individuals in the age group 60-70  at $t_d^2$ were 40-50 years old at time $t_b^1=5$.  Thus,this cohort was impacted by adverse swap events. On the other hand, individuals of age 60-70 at time $30$ were 35-45 years old at  $t_b^1$, which means that individuals under 40 at this time were not impacted by adverse swap events. This cohort is thus less deprived than the older cohort. This is also the case for older age groups, but the impact is delayed. 

\begin{figure}[H]
\hspace{-1.7cm}
\begin{subfigure}[t]{.6\textwidth}
\centering
\includegraphics[scale=0.42]{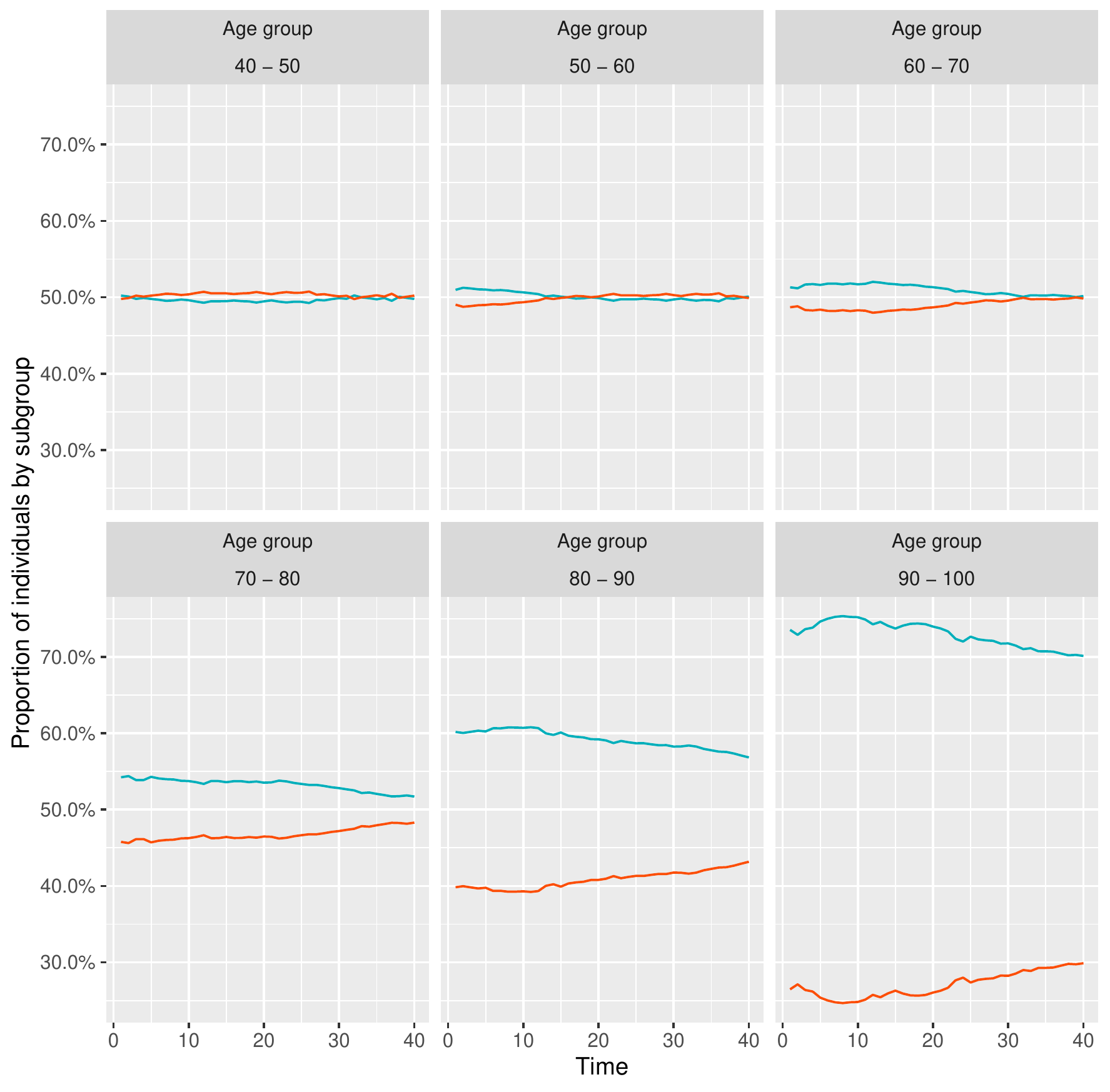}
\caption{ Model 1 (no swaps).}
\label{compoagegrpnoswap}
\end{subfigure}
\begin{subfigure}[t]{.6\textwidth}
\centering
\includegraphics[scale=0.42]{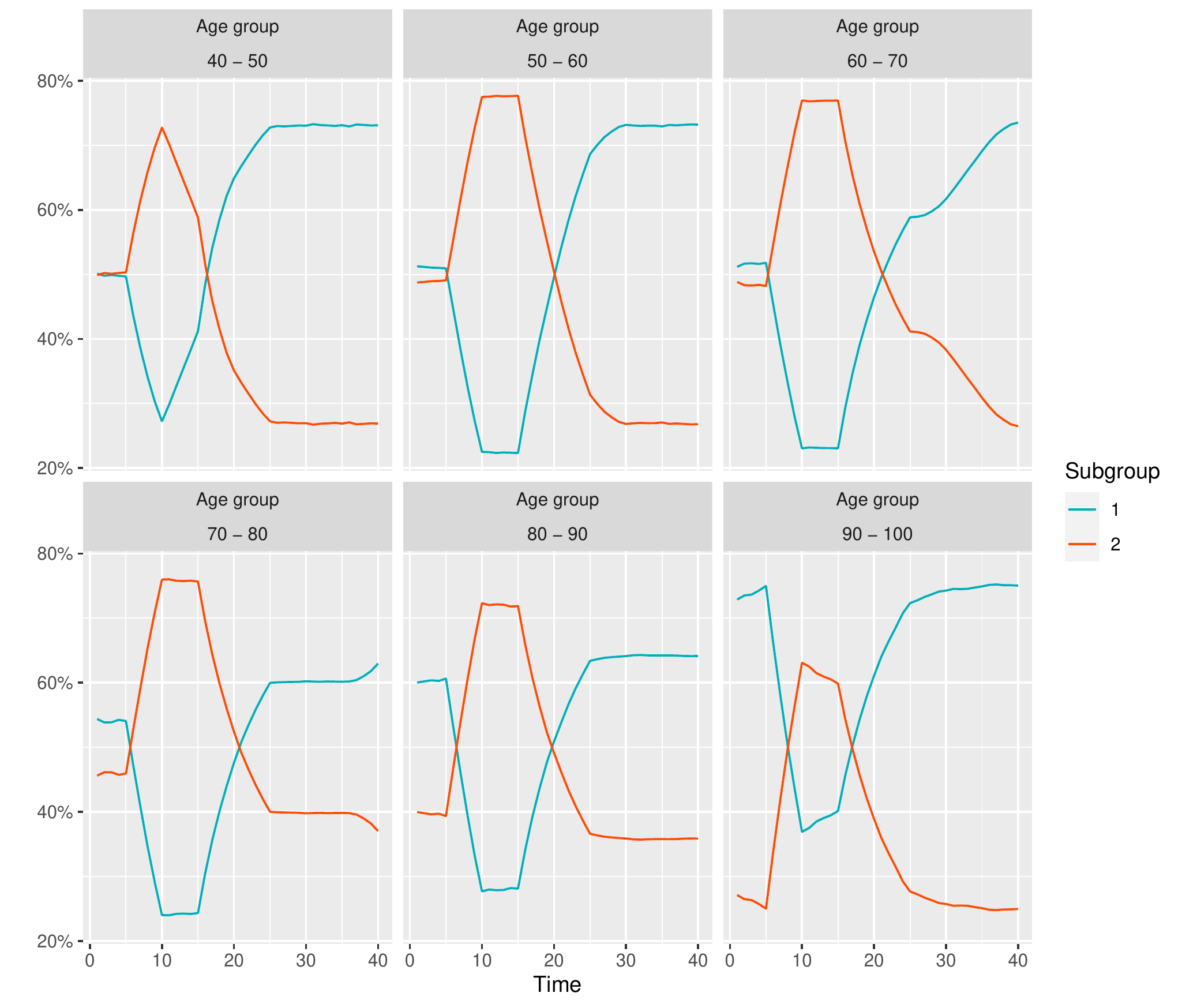}
\caption{ Model 2 (with swaps).}
\label{compoagegrpswap}
\end{subfigure}
\caption{Evolution of age groups composition over time.}
\label{compoagrgrp}
\end{figure}

\noindent  \textbf{Lee-Carter model} These nonlinear changes in the population composition, which differ significantly across time and age groups, have varying effect on the general population's mortality rates. In both models, we fit a Lee-Carter model to the simulated data in both models, using the R package  StMoMo. First, the mortality shock over the period $[t_b,t_d[$ is actually not well-captured by the Lee-Carter model, even when the model is only fitted for ages above 65. This is due to the fact that the magnitude of the shock increases with age (which is case for  the COVID-19 pandemics), while the parameters $(k_t)$ in the Lee-Carter model only allows a fixed effect for all ages, and the parameters $(\beta_x)$ determine the relative changes in mortality of each age class, but for the whole period and not only during the mortality shock. This raises the question of the ability of the Lee-Carter model to to estimate mortality rates in the presence of abrupt changes, which is even more complicated when the population composition is modified. \\
The model implemented are stochastic and there is an inherent demographic noise (or idiosyncratic risk) in the estimated central death rates.  In order to compare mortality rates in both models, we have fitted  two Lee-Carter models over the 40 year period of the simulation: one for the age period 55-79 (on year age classes) and one for the age period 80-95. Figure \ref{LCdeathrates} shows the evolution the estimated log-mortality rates in both models, at ages 55, 65, 75 and 85. In both models, we can observe the first impact of the mortality shock between year 2 and 5, which increase with age.  Swap events (brown curve) from year 5 to year 10 have a strong impact in the population mortality. The adverse compositional changes following the mortality shock not only compensate the decline in mortality rates in each subgroup but also increase the general population. This increase is also more important when age increase.  From year 15, the long-term "positive" effect of the mortality generates modification in the population which accelerate the mortality decline, and mortality rates become actually in the second model than in the first model at time 40. 

\begin{figure}[H]
  \begin{subfigure}[t]{.4\textwidth}
    \centering
    \includegraphics[width=\linewidth]{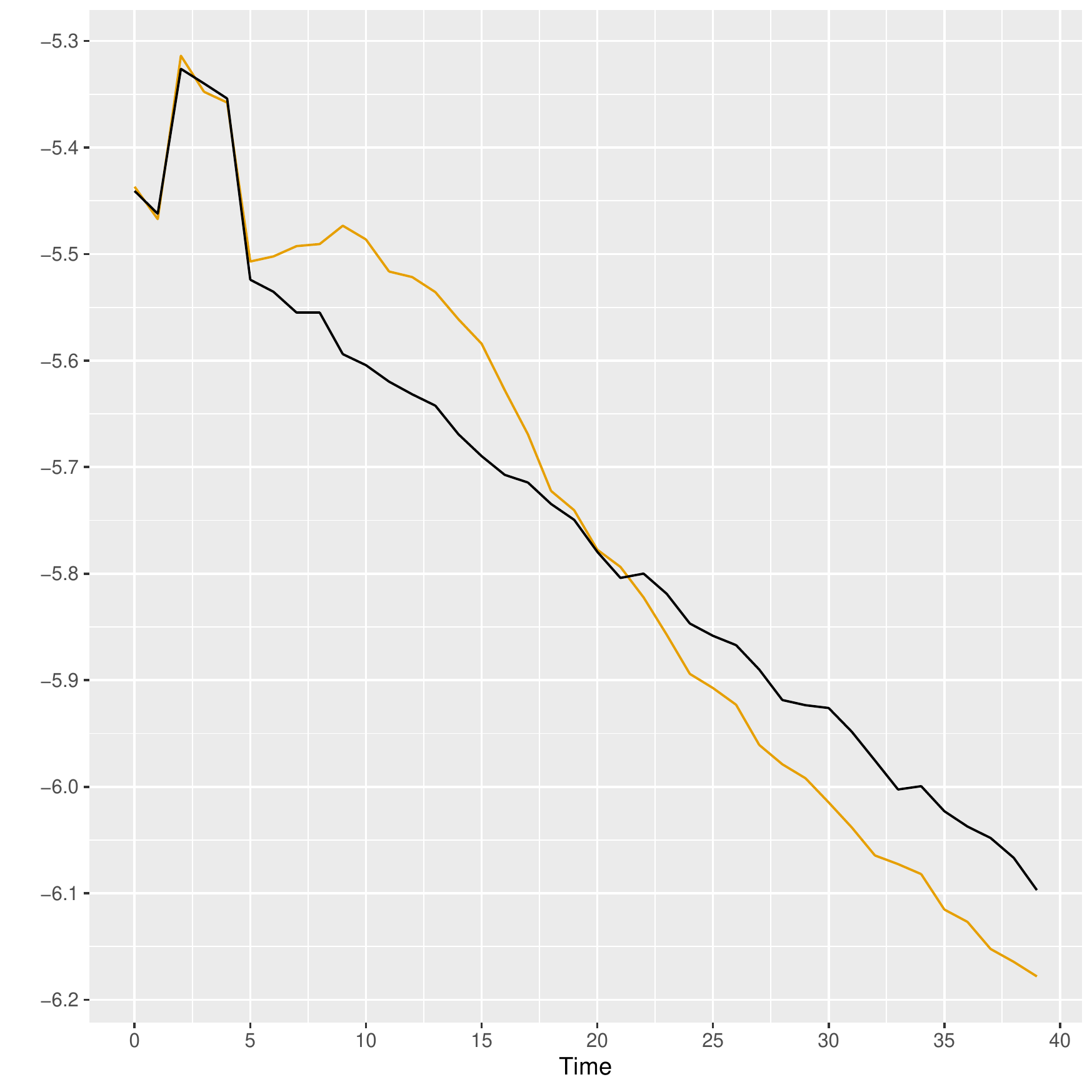}
    \caption{Age 55}
  \end{subfigure}
  \hfill
  \begin{subfigure}[t]{.4\textwidth}
    \centering
    \includegraphics[width=\linewidth]{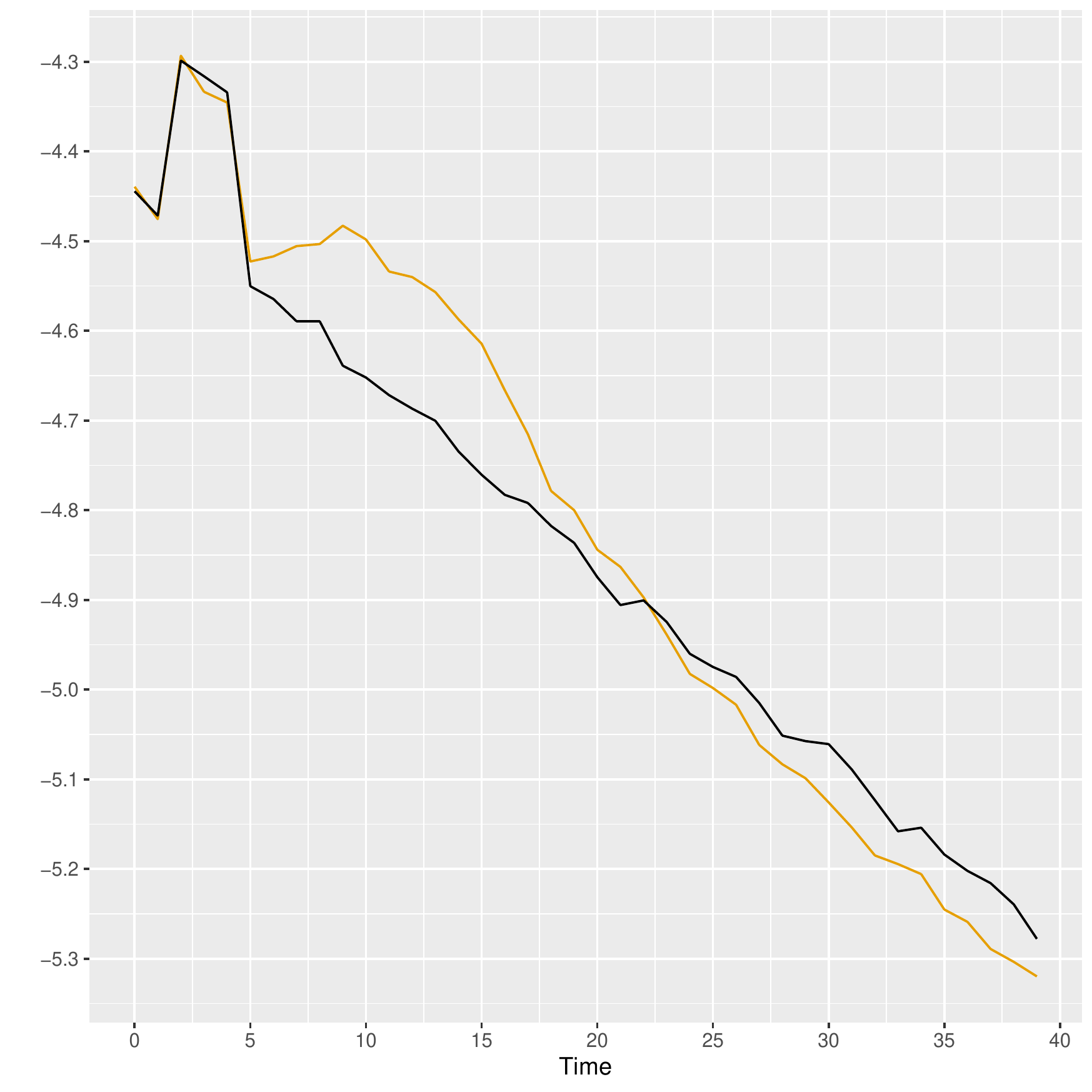}
    \caption{Age 65}
  \end{subfigure}

  \medskip

  \begin{subfigure}[t]{.4\textwidth}
    \centering
    \includegraphics[width=\linewidth]{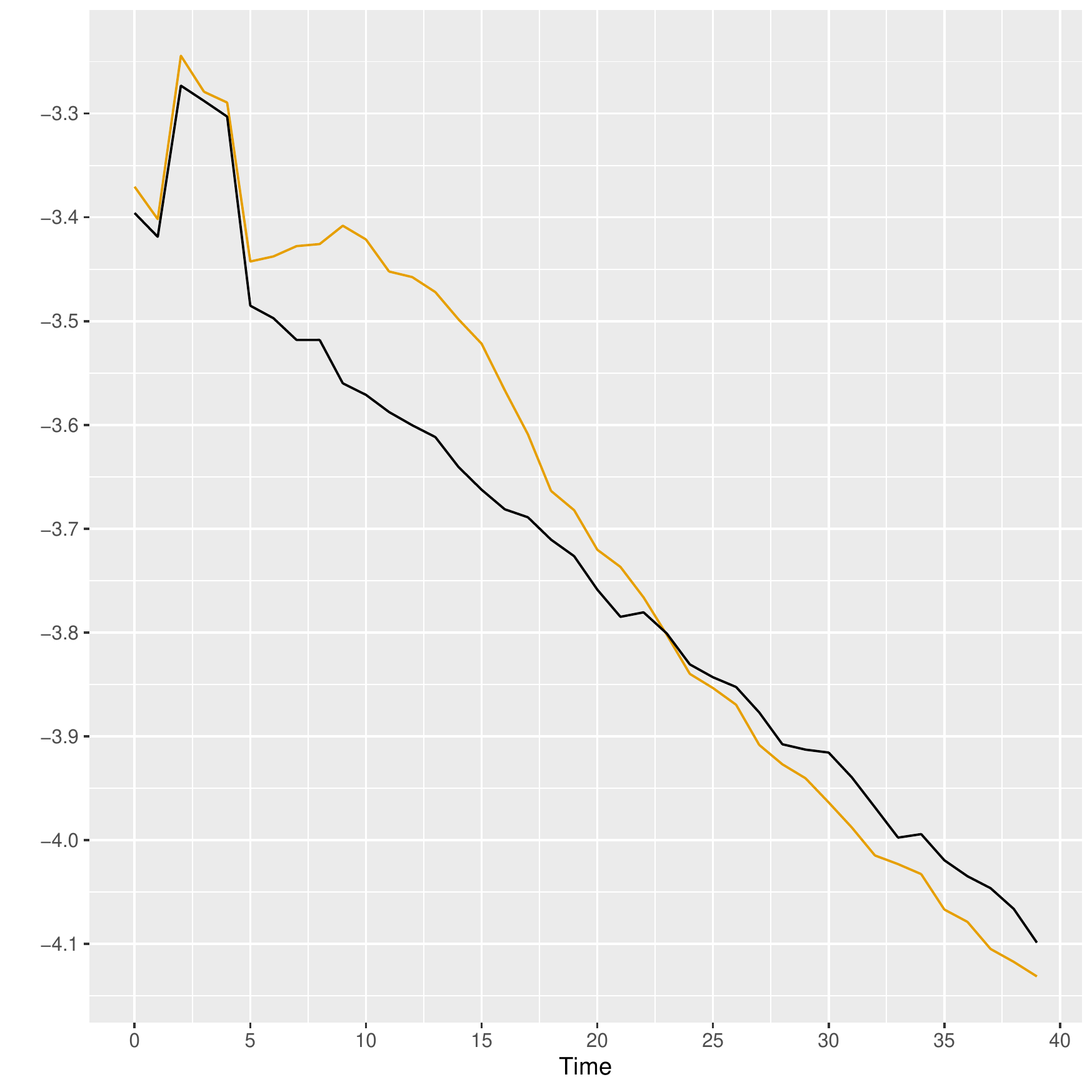}
    \caption{Age 75}
  \end{subfigure}
  \hfill
  \begin{subfigure}[t]{.4\textwidth}
    \centering
    \includegraphics[width=\linewidth]{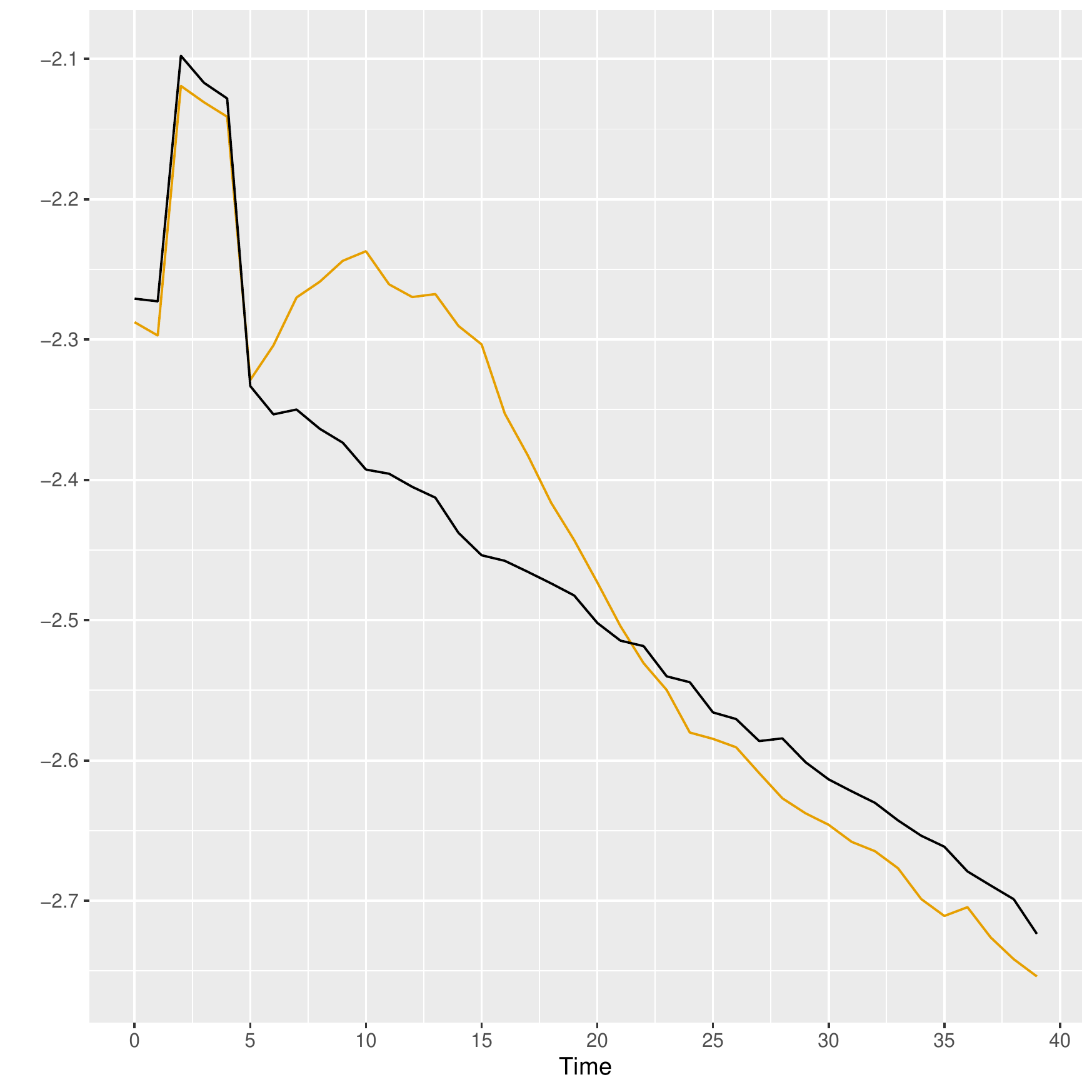}
    \caption{Age 85}
  \end{subfigure}
  \caption{Estimated  log mortality rates (Lee-Carter)- Model with Swaps (brown) and with no Swaps (black).}
  \label{LCdeathrates}
\end{figure}

\subsection{Conclusion and perspectives} These experimentation show the potential major long-term impact on longevity of interactions and changes in the population composition, for instance due to macro public policies. Interpreting longevity trends  in an heterogeneous population is an arduous task, especially when only analyzing aggregated data. The forecasting of mortality rates, following abrupt changes which could generate social changes and redistribution of the population composition is even more difficult. Even in the simplified model presented in this section, the standard forecast of mortality rates from years following the mortality shock would yield substantial errors, due to the non-stationary 
and opposite mechanisms underlying the population dynamics.\\
 This raises many questions regarding the forecast of mortality rates following the COVID-19 pandemics, in light of the many long term public health issues that we are currently facing.  If taking into account population data as a complement to mortality data appears  important, if not necessary to understand future trends, the modeling of the population dynamics remains a complex task. Determining relevant subgroups can be a challenge, although several measures of the general socioeconomic status of individuals or geographically based index of deprivation have been recently proposed and integrated in mortality models (see e.g. \cite{villegas2014modeling}, \cite{shang2017grouped}). Estimating swap rates when there are more than two subgroups or in an open population is also difficult, often due to lack of individual data. A partial answer would be to include indicators of the population composition as explanatory variables in statistical mortality models for the general population. \\
The flexible  framework of Individual-Based-Models is also an important tool for experimenting with various demographic scenarios and integrate interdisciplinary information, especially in the current period of high uncertainty regarding future trends in longevity. The difficulty posed by high computational costs of Individual-Based Models have been overcome by efficient algorithms implemented in IBMPopSim (simulations runtimes for the models presented in this paper are less than 2 minutes).  In the toy model presented in this paper, the impact of public policies or changes in the global socioeconomic or environmental context are modeled by changes in the population, but further modeling could also include macro-variables as explanatory variables of demographic rates.

\bibliographystyle{apalike}
{\footnotesize


\bibliography{biblioSurveyalphalon2017Juillet2017}
}
\end{document}